\documentclass[11pt]{article}

\usepackage{graphicx}
\usepackage{bm}

\begin{document}

\title{\bf COMMENTS ON CAHILL'S QUANTUM FOAM INFLOW THEORY OF GRAVITY}         
\author{T. D. Martin}        
\date{Gravity Research Institute \linebreak Boulder, CO 80306-1258 USA \linebreak martin@gravityresearch.org}          
\maketitle

\abstract{\em{We reveal an underlying flaw in Reginald T. Cahill's recently promoted quantum foam inflow theory of gravity.  It appears to arise from a confusion of the idea of the Galilean invariance of the acceleration of an individual flow with what is obtained as an acceleration when a homogeneous flow is superposed with an inhomogeneous flow.  We also point out that the General Relativistic covering theory he creates by substituting a generalized Painlev\'{e}-Gullstrand metric into Einstein's field equations leads to absurd results.}}
\linebreak

\section{Introduction}       

Recently, Reginald T. Cahill has been promulgating a theory of gravity based on the idea of a quantum foam which is envisioned as flowing into gravitational attractors.  The dynamics of the flow is determined by a classical Euler-type fluid flow equation.   At the same time, he has been asserting that data from historical Michelson-Morley-type experiments performed on the surface of the Earth show evidence of a solar gravitational inflow and a flow arising from the absolute motion of the solar system through the cosmos (see \cite{1,5,6} and further references listed therein).  In order to make the idea of a gravitational inflow fit together with the idea of a detectable absolute motion of the solar system, he claims to be able to vectorially superpose the Earth's inhomogeneous gravitational inflow with the quasi-homogeneous background flow and not change the gravitational attraction of the overall flow.  In support of this claim, he thinks that he has provided a proof of the assertion that the inhomogeneity of the vector sum of an inhomogeneous flow and a homogeneous flow does not depend on the homogeneous flow.  In the next Section, we will quickly disprove these latter claims and show how they lead to absurd results from a physical point of view.  In Section 4 we will alert the reader to the unacceptable physical predicitions which arise from Cahill's substitution of the Painlev\'{e}-Gullstrand metric into Einstein's field equations as a means of creating a sort of General Relativistic foundation for his quantum foam inflow theory.  By uncovering these flaws, we hope to protect the reputations of future theories of gravity involving some sort of "flowing substratum" or, what is more likely, those involving some sort of "preferred frame flow".   

\section{The Underlying Flaw}

Very quickly, we refer to Cahill's preprint \cite{1} as being typical of his series of papers on his inflow theory.  Proceed to his fundamentally important equation (48),

\begin{equation}
{\bf g} = {1 \over 2}\nabla ({\bf v}^2 ) + {{\partial {\bf v}} \over {\partial t}}
\end{equation}which is purported to give the gravitational acceleration arising from an irrotational substratum flow ${\bf v}$ (the flow of the quantum foam).  We will apply this equation to the approximately time-independent superposition of the Earth's supposed gravitational inflow (Cahill's equation (6))

\begin{equation}
{\bf v} \equiv  - \sqrt {{{2GM} \over r}} \,\,{\bf \hat r}
\end{equation}and the approximately time-independent and homogeneous motion of the substratum relative to the Earth

\begin{equation}
{\bf V} \cong {\bf const}.
\end{equation}${\bf V}$ is the superposition of the solar gravitational inflow, the flow induced by the Earth's orbital motion, and the flow induced by the absolute motion of the solar system through the regional substratum all summed at the nexus of the surface of the Earth.  Its magnitude will always be roughly 400 km/sec, because the speed of the solar inflow is about 42 km/sec, the speed of the Earth's orbital motion is about 30 km/sec, and, according to Cahill, the speed of the absolute motion of the solar system is  
$417 \pm 40$ km/sec (see page 19 of his preprint \cite{1}).

Using the vector identity

\begin{equation}
\nabla ({\bf v} \cdot {\bf V}) = ({\bf v} \cdot \nabla ){\bf V} + {\bf v} \times (\nabla  \times {\bf V}) + ({\bf V} \cdot \nabla ){\bf v} + {\bf V} \times (\nabla  \times {\bf v})
\end{equation}we calculate Cahill's prediction (our equation (1)) for the supposed gravitational field around the Earth arising from the approximately time-independent superpostion ${\bf v} + {\bf V}$ as follows.  Since ${\bf V}$ is a constant and all curls are zero,
$${\bf g} = {1 \over 2}\nabla (({\bf v} + {\bf V})^2 ) = {1 \over 2}\nabla (({\bf v} + {\bf V}) \cdot ({\bf v} + {\bf V})) = {1 \over 2}\nabla ({\bf v} \cdot {\bf v} + 2({\bf v} \cdot {\bf V}) + {\bf V} \cdot {\bf V})
$$

$$
 = {1 \over 2}\nabla (v^2 ) + \nabla ({\bf v} \cdot {\bf V}) + {1 \over 2}\nabla (V^2 )
$$

$$
 =  - {{GM} \over {r^2 }}\,{\bf \hat r} + ({\bf V} \cdot \nabla ){\bf v} + ({\bf v} \cdot \nabla ){\bf V} + {\bf 0}
$$By definition,

\begin{equation}
({\bf V} \cdot \nabla ){\bf v} \equiv {\bf V} \cdot (\nabla {\bf v})
\end{equation}and the tensor (dyadic) $\nabla {\bf v}$ is (cf. reference \cite{2})

\begin{equation}
\nabla {\bf v} = \sqrt {{{GM} \over {2r^3 }}} {\bf \hat r}\,{\bf \hat r} - \sqrt {{{2GM} \over {r^3 }}} \hat {\bm \theta} \,\hat {\bm \theta}  - \sqrt {{{2GM} \over {r^3 }}} \hat {\bm \phi}\,\hat {\bm \phi} 
\end{equation}Since the tensor $\nabla {\bf V}$ is zero, we have

\begin{equation}
{\bf g} =  - {{GM} \over {r^2 }}{\bf \hat r} + ({\bf V} \cdot {\bf \hat r})\sqrt {{{GM} \over {2r^3 }}} {\bf \hat r} - ({\bf V} \cdot \hat {\bm \theta} )\sqrt {{{2GM} \over {r^3 }}} \hat {\bm \theta}  - ({\bf V} \cdot \hat {\bm \phi} )\sqrt {{{2GM} \over {r^3 }}} \hat {\bm \phi} 
\end{equation}

This equation gives absurd results for the gravitational field at the surface of the Earth.  As we have said, the solar-system-flow plus orbital-flow plus solar-gravitational-flow {\bf V} will be an approximately homogeneous flow when compared to the terrestrial gravitational influx {\bf v}.  At some point on the surface of the Earth, this homogeneous flow and the terrestrial influx will be maximally vectorially summed (call this the "forward" position), and at the antipodal point of the Earth, this sum will be minimized (call this the "aft" position).  If one substitutes the physical constants \emph{G}, \emph{M}, and \emph{r} which are appropriate for the surface of the Earth into equation (7) along with the magnitude of {\bf V} being taken as 400 km/sec, one discovers that the equation predicts a huge gravitational attraction on the forward side of the homogeneous flow at the surface of the Earth (actually, 36.1 times greater than the normal 9.82 ${\rm m}{\rm /}{\rm sec}^{\rm 2}$ value) and a huge \emph{repulsion} on the aft side of the flow at the surface of the Earth (with a magnitude which is 34.1 times greater than the normal 9.82 ${\rm m}{\rm /}{\rm sec}^{\rm 2}$ value).

There is a very simple way to understand the essence of these results concerning the superposition of the homogeneous background flow and the Earth's gravitational inflow.  At the forward position on the surface of the Earth, the speed of the superposed flow is increasing in the direction towards the Earth's center (because the background flow and the inflow are in the same direction).  Since the gravitational acceleration is given by the gradient of the square of the speed of the flow, there is a gravitational attraction at the forward position.  In contrast, at the aft position on the surface of the Earth, the speed of the superposed flow is increasing in the direction away from the Earth's center (because the background flow and the inflow are oppositely directed).  Thus, there will be a gravitational repulsion at the aft position.

\section{Discussion}

Cahill believes that he has detected an absolute motion of the solar system in the historical data of various terrestrial Michelson-Morley-type experiments.  His interpretation of the data implies that there is a flow everywhere on the surface of the Earth and that the magnitude of the flow is always on the order of 400 km/sec.  Cahill also believes that the Earth acts as a sink and that there is a corresponding terrestrial inflow.  In order to allow himself to superpose the quasi-homogeneous cosmic flow with the Earth's inhomogeneous gravitational inflow and still retain ordinary Newtonian attraction, he erroneously claims to demonstrate that the acceleration field {\bf g} resulting from the superposition of an inhomogeneous flow and a homogeneous flow is the same as the acceleration field of the inhomogeneous flow alone (see page 17 of \cite{1}).  That his argument is misleading is mathematically proven by the result we have obtained in equation (7) above.  The fact that the acceleration of any particular flow governed by Euler's equation is Galilean invariant cannot be taken to imply that the acceleration of the superposition of an inhomogeneous flow with a homogeneous flow is the same as the acceleration of the inhomogeneous flow alone.  The acceleration associated with a sink which is sitting at rest in its background fluid (i.e., a sink which is not moving relative to the fluid very far away from the sink) is certainly Galilean invariant, but it is not the same as the acceleration associated with the flow that is produced when the sink itself is moving with respect to the background fluid.

Cahill faces a logical conundrum. On the one hand, if there really are terrestrial and solar gravitational substratum flows, then, because one cannot vectorially sum the flows and retain Newtonian gravity near their separate sources (as we have shown above), the flows are necessarily entrained and separated by a gravitopause as we have explained in the references \cite{3,4}.  In that case, the solar gravitational flow and the absolute motion of the solar system would not show up in the historical Michelson-Morley-type data.  On the other hand, if Cahill has truly discovered a real physical cosmic flow in the old Michelson-Morley data, he cannot have the gravitational inflow he desires in his theory.  For this reason, it seems unlikely that real solar or terrestrial inflows would also be seen in that data.

\section{The Painlev\'{e}-Gullstrand Substitution}

Since Cahill continues to promote and develop his theory \cite{5}, we think it is important for people to know that his theory has other problems besides the superposition problem.  As just one example, Cahill seems to have started talking about his quantum foam as an \emph{inflow} about the time he cited one of our papers \cite{3} in one of his own \cite{6}.  We get the impression that Cahill jumped the gun a little bit by taking a tentative suggestion we made on page 23 of our paper \cite{3} at face value.  This was about the possibility of substituting the flow metric (a generalized "Painlev\'{e}-Gullstrand" metric) into Einstein's field equations.  In fact, Cahill went ahead and did this, and he has used the resulting equations as a major component of his theory ever since (see, for example, Section 4 of his preprint \cite{5}).
  
There were several cogent reasons for our not having substituted the Painlev\'{e}-Gullstrand metric into Einstein's field equations in our paper \cite{3}, not the least of which was that the Painlev\'{e}-Gullstrand metric has only three free parameters, while it is common knowledge that a generally applicable coordinate system (i.e., a generally applicable space-time "chart") in General Relativity must have six such parameters \cite{7}.  However, there is a much more important reason for not basing a general theory of an inflowing quantum foam on the undiscerning substitution of the Painlev\'{e}-Gullstrand metric into Einstein's field equations.  As we carefully pointed out in a subsequent paper \cite{4}, this substitution has the absolutely necessary consequence that, in the interior of any spherically symmetric and stationary mass-energy distribution, the radial stress of the mass-energy distribution is everywhere \emph{equal} to its mass-energy density (see especially Section 2 of reference \cite{4}).  This can only be true for certain electromagnetic energy distributions and for extraordinary states of matter.  In other words, as a consequence of his substituting the Painlev\'{e}-Gullstrand metric into Einstein's field equations, Cahill's theory gives absurd predictions for the interior states of matter in all ordinary planets.


\begin{thebibliography}{9}

\bibitem{1}R. T. Cahill, \textit{Gravity as Quantum Foam In-flow}, physics/0307003

\bibitem{2}R. S. Brodkey, \textit{The Phenomena of Fluid Motion}, Dover Publications, N.Y. (1995), p. 49

\bibitem{3} T. D. Martin, \textit{General Relativity and Spatial Flows: I. Absolute Relativistic Dynamics}, gr-qc/0006029  

\bibitem{4} T. D. Martin, \textit{General Relativity and Spatial Flows: II. The Hollow Shell Cavendish Experiment}, http://www.gravityresearch.org/pdf/GRI-010515.pdf

\bibitem{5} R. T. Cahill, \textit{'Dark Matter' as a Quantum Foam In-flow Effect}, physics/0405147

\bibitem{6} R. T. Cahill, \textit{Process Physics: From Quantum Foam to General Relativity},
gr-qc/0203015

\bibitem{7} S. Weinberg, \textit{Gravitation and Cosmology}, John Wiley and Sons, N.Y. (1972), p. 161 	


\end{thebibliography}
\end{document}